\documentclass[10pt,leqno]{article}

\usepackage{cite}
\usepackage{amsmath,amssymb,amsfonts}
\usepackage{algorithmic}
\usepackage{graphicx}
\usepackage{textcomp}
\usepackage{xcolor}
\usepackage{caption}

\usepackage{graphicx}
\usepackage{indentfirst,csquotes,authblk}
\usepackage{xcolor}

\usepackage{amssymb,amsthm,amsmath}
\usepackage{xcolor,paralist,hyperref,titlesec,fancyhdr,etoolbox}

\titleformat{\section}[block]{\normalfont\Large\bfseries\centering}{\centering\thesection}{10pt}{\Large}
\titlespacing*{\section}{0pt}{0ex}{0ex}

\hypersetup{ colorlinks=true, linkcolor=black, filecolor=black, urlcolor=black }

\title{RISC-V V Vector Extension (RVV) with reduced number of vector registers} 
 
\author[1] { Eino  Jacobs \thanks{eino.jacobs@synopsys.com }}
\author[1] { Dmitry  Utyansky \thanks{dmitry.utyansky@synopsys.com}}
\author[2] { Muhammad Hassan \thanks{Muhammad.Hassan@infineon.com}}
\author[2] { Thomas Roecker \thanks{Thomas.Roecker@infineon.com}}
\affil[1] { Synopsys, Inc, Sunnyvale, USA }
\affil[2] { Infineon Technologies AG, Munich, Germany }

\begin{document}

\maketitle

\let\thefootnote\relax

\begin{abstract}
To reduce the area of the RISC-V Vector extension (RVV) in small processors, the authors are considering one simple modification: reduce the number of registers in the vector register file. The standard “V” extension requires 32 vector registers that we propose to reduce to 16 or 8 registers.  Other features of RVV are still supported.

Reducing the number of vector registers does not generate a completely new programming model: although the resulting core does not have binary code compatibility with the standard RVV, compiling for it just requires parameterization of the vector register file size in the compiler.

The reduced vector register file still allows for high utilization of the RVV vector processor core. Many useful signal processing kernels require few registers.
 
\end{abstract} 

\bigskip

\section{Introduction}

The RISC-V Vector Extension (RVV), ratified in version 1.0 \cite{rvv10}, provides an Instruction Set Architecture (ISA) that has scalability and flexibility with respect to the underlying processor implementation. This allows optimization of processor designs for different area/power/performance targets that all share a consistent programming model. One of the advantages of this approach is cost reduction through reuse of hardware, compilers, libraries and software.

However, for low-end processors the mandatory vector register file with 32 vector registers contributes visibly to the overall core size. In this paper we assume that the smallest practical vector length ($VLEN$) is 64 bits and therefore the smallest practical vector register file consists of 32 registers of 64 bits for a total of 2048 bits. We also assume that the number of physical vector registers equals the number of architectural vector registers. The RVV standard allows for a VLEN that is lower than 64, VLEN of 32 bits, but we do not consider this practical, because ecosystem support is currently weak for this small size and 64-bit data types are not supported. We argue that an option to reduce the vector register file down to 16 or even 8 registers is a good choice for some designs. A vector register file with 8 registers of 64 bits has in total 512 register bits. This is the same number of bits as in a scalar register file with 16 registers of 32 bits that are in an RV32E version of RISC-V.

\section {Number of Vector Registers, their Sizes and Typical Application Requirements}

The number of vector registers and the length of vector registers must be considered together for performance analysis. Depending on the processor data path length ($DLEN$), available functional units and other details, the dimensions of the vector register file that are required to obtain good cycle performance and good hardware utilization will vary. This also depends on compute kernels: complexity of computations, our ability to reuse data loaded into registers, etc., and, therefore, depends on the application domain. In this paper we focus on embedded applications and processor configurations minimizing processor size and power consumption. At this area/performance point the processor has a gate count of a few dozen kgates, datapath length of 32 bits ($DLEN=32$), and either in-order single-issue or limited fusion/multi-issue capabilities. 

One attractive feature of RVV is potential "quasi-multi-issue" through chaining. If the vector length is longer than DLEN, a vector instruction issued in one cycle is executed for multiple cycles ($ VLEN/DLEN > 1 $ cycle). The extra cycles can be used to issue other instructions. These can be executed in parallel with the first vector instruction (in the same \textit{convoy} using vector processors terminology), as long as they use other execution resources. One can think of \textit{chaining ratio} as $(EMUL \cdot VLEN)/DLEN $, where $EMUL$, or effective $LMUL$ (vector length multiplier), is defined by RVV 1.0 \cite{rvv10} as the number of registers required to hold vector operand elements. So, with e.g. chaining ratio of 1:4 one can see the code as packages (convoys) of 4 multicycle instructions using different execution resources.

A few considerations need to be taken into account to dimension the vector register file. 
\begin{itemize}
\item Cycle performance and the program's ability to fully load the processor's functional units. There is a sweet spot for the chaining ratio, beyond which the execution resources are fully saturated, therefore for a given DLEN there is no benefit in increasing VLEN beyond that.
\item Processor size. For small processors, the vector register file is a substantial contributor to the area, so obviously the smaller the better.
\item Applications. There is a limit to vector length that the application can use. For typical embedded applications this is naturally not too big.
\end{itemize}

We observe that in small processor designs the full RVV 1.0 vector register file (32 64-bit vector registers) can take about 30\% of the overall logic area. In this case, a reduction of the vector register file by half (16 64-bit vector registers) or to a quarter size (8 64-bit vector registers) can save 15\% to 23\% of the processor area, respectively. 

For application examples, consider the automotive industry. One such application is the filtering of ADC data, necessitating the use of algorithms such as Finite Impulse Response (FIR) or averaging filters. Another application is the enhancement of sensor data accuracy through virtual sensors or filtering of sensor data. This typically involves system state modelling with Kalman Filters or  Embedded AI Neural Network (NN) topologies like Multilayer Perceptrons (MLP), autoencoders and Recurrent Neural Networks (RNNs). Additionally, audio sensor data processing, such as key word recognition and environment sensing using audio, is another area for which embedded processors are utilized. These algorithms typically rely heavily on matrix and vector computations. Efficient execution of these algorithms on embedded processors is crucial for ensuring the reliability and accuracy of the automotive systems that rely on them.

Compute kernels used in linear algebra and Digital Signal Processing (DSP) usually do not require huge amounts of data in flight. E.g., a simple "load-load-multiply-accumulate" pattern can be implemented with just the bare minimum of 2 vector registers for input, 1 vector register for output. 

RVV-like vector architectures allow to avoid explicit software pipelining and loop unrolling, with multiple loop iterations in flight, which would otherwise require multiple copies of data in registers, requiring more registers. Essentially hardware takes care of the required pipelining, enabling parallelism of e.g. loads and ALU operations. 

Many well-known vector processors used fairly few vector registers with good results. E.g., all Cray computers had 8 vector registers (\cite{cray-1} \cite{cray-y}), though of longer length each. Closer to modern days, ARM M-profile Vector Extensions (MVE) use 8 128-bit vector registers \cite{arm_m55}. RVV's 32 vector registers together with the ability to combine vectors into groups using LMUL factor allows to configure the processor for fewer longer vector registers: with LMUL=8 one effectively has 4 vector registers of 8x longer size. 

\section { Processor Configuration for Embedded DSP }

In the subsequent sections we focus on fixed point kernels for small embedded DSP use-cases. We assume a small processor without floating point; inclusion of a relatively large floating point unit would make the area savings of a reduction of the number of vector registers seem less important.

Other features we assume:
\begin{itemize}
\item $VLEN=64$ bits, floating point deconfigured (i.e., Zve64x extension). 64 bits as minimum also makes sense because 64-bit integers are useful as accumulators.
\item $DLEN=32$ bits in the smallest configuration. This can be scaled up for bigger processors, with the matching vector length increase.
\item Processor implementation supporting chaining of operations (at least between vector load and ALU and Multiply-accumulate unit).
\item No other multi-issue capabilities. Chaining is the only way to organize "quasi-multi-issue".
\item Load/store bandwidth of 32 bits. A higher bandwidth of 64 bits ($DLEN\cdot 2$) would relax the load bottleneck for some kernels, This is a topic to explore further. 
\item Reasonably short latency of load, MAC and ALU operations. Typically such designs use relatively low clock frequency, hence shallow processor pipeline. For the examples considered we use 1-cycle latency for loads, 5 cycles for multiply-accumulate, 3 cycles for other ALU operations. For the kernels considered in this paper a small variation of the latency does not change the results.

\end{itemize}

\section {DSP Kernel Analysis with Reduced Vector Registers }

In this chapter we analyze several digital signal processing kernels, useful for many applications. To quantify the quality of mapping of the kernel to the processor we use utilization of the critical resource for each kernel. Typically this is either ALU, multiplier, or load/store bandwidth.

One important consideration is "run-time $VLEN$ agnosticism". Depending on the use case, this might be more or less important. For deeply embedded applications it often does not matter: the program is compiled for the specific processor, and knowledge about its parameters, like hardware vector length, can be used by the compiler to better optimize the code. Specifically that allows to drop "vset*" instructions from tight loops, minimizing number of instructions. Also, special tail handling often can be avoided by choosing convenient sizes or explicit tail handling outside of the main loop. In the subsequent analysis we assume that a program is compiled for the specific configuration, so vset* instructions can be dropped from the inner loops.

\subsection { Matrix Multiplication }

Each table in this and the subsequent sections shows the inner kernel of the processing loop in sustained operation, with columns showing the instructions issued at each cycle, the functional unit used by the instruction, and the results written to the vector register file by the selected functional units at that cycle. A colon-separated number after vector register name is a sequential number of $DLEN$-sized group of bits output at that cycle. For ($VLEN=64$, $DLEN=32$)-processor each vector register is treated as two such groups, :0 and :1. So e.g. writes of v0:0,v0:1 as a result of vwmacc v0  operation appear +5 cycles later, while the preceding writes by the MAC correspond to the previous loop iteration. Vmacc is a widening multiply-accumulate operation, so it generates 2x wider result, two $DLEN$-sized chunks.

For the reference processor configuration, starting with the tile size of $2\times(VLEN\cdot2)$ (e.g. $2\times16$ samples for 8-bit samples and $VLEN=64$) we can get to 8/9 = 89\% utilization of the multiply-accumulate unit. For this we need 1:4 chaining ratio, achievable with LMUL=2, as shown in Table \ref{GEMM_2xVLEN_m2}. Total of 9 cycles in the loop (assuming single issue, one instruction per cycle, no stalls), of which MAC unit is active for 8 cycles, producing double-wide result in each cycle.

For some cases the utilization can be 100\%. If one of the scalar loads can use immediate offset, both loads can reuse the same base address and so only one address increment is required, reducing number of instructions to 8, as shown in Table \ref{GEMM_2xVLEN_b_m2}.

If $2\times16$ tiles are too wide for the application, inevitably we have fewer data processed with the same code and same cycles, so e.g. for $2\times 8$ tiles we will have LMUL=1 and utilization of 4/9=44\%, as shown in Table \ref{GEMM_2xVLEN_m1}. Or 50\% if just one address update \textit{addi} is needed, i.e., when overall matrix row size fits into a 12-bit immediate offset allowed in the \textit{lb} instruction.

For comparison, increasing the tile width to fit LMUL=4 allows us to obtain 100\% utilization, with enough "issue cycles" to spare that can be potentially used for some additional computations while maintaining the same MAC utilization, as shown in Table \ref{GEMM_2xVLEN_m4}

\begin{center}
\small
\texttt{
        \begin{tabular}{ |l | l | c | c | c |}
        \hline
Cycle & Instruction issued  &  Unit &       VMAC:M      &  VLOAD:L \\
        \hline
0 & vle8.v         v8,(x28)               &     VLOAD & M:v1:0,v1:1  &  L:v8:0 \\
1 & c.addi         x28,0x10               &    Scalar & M:v2:0,v2:1  &  L:v8:1 \\
2 & lb             x7,(x13)               &    Scalar & M:v3:0,v3:1  &  L:v9:0 \\
3 & vwmacc.vx      v0,x7,v8              &      VMAC & M:v4:0,v4:1  &  L:v9:1 \\
4 & c.addi         x13,0x1                &    Scalar & M:v5:0,v5:1  &       .  \\
5 & lb             x8,(x5)                &    Scalar & M:v6:0,v6:1  &       .  \\
6 & c.addi         x5,0x1                 &    Scalar & M:v7:0,v7:1  &       .  \\
7 & vwmacc.vx      v4,x8,v8              &      VMAC &      .       &       .  \\
8 & bne            x13,x15,0xffffffe6     &    Scalar & M:v0:0,v0:1  &       .  \\
       \hline
        \end{tabular}
\captionof{table}{Matrix multiplication kernel, $2\times VLEN\cdot 2$ tile, $8*8 \rightarrow 16$ bits \\10 vector registers used}\label{GEMM_2xVLEN_m2}
        }
    \end{center}

\begin{center}
\small
\texttt{
        \begin{tabular}{ |l | l | c | c | c |}
        \hline
Cycle & Instruction issued  & Unit  &       VMAC:M      &  VLOAD:L \\
        \hline
0 & vle8.v         v8,(x28)               &     VLOAD & M:v1:0,v1:1  &  L:v8:0 \\
1 & c.addi         x28,0x10               &    Scalar & M:v2:0,v2:1  &  L:v8:1 \\
2 & lb             x7,N(x5)               &    Scalar & M:v3:0,v3:1  &  L:v9:0 \\
3 & vwmacc.vx      v0,x7,v8              &      VMAC & M:v4:0,v4:1  &  L:v9:1 \\
4 & lb             x8,(x5)                &    Scalar & M:v5:0,v5:1  &      .  \\
5 & c.addi         x5,0x1                 &    Scalar & M:v6:0,v6:1  &       .  \\
6 & vwmacc.vx      v4,x8,v8              &      VMAC &  M:v7:0,v7:1  &       .  \\
7 & bne            x13,x15,0xffffffe6     &    Scalar & M:v0:0,v0:1  &       .  \\
       \hline
        \end{tabular}
\captionof{table}{Matrix multiplication kernel, $2\times VLEN\cdot 2$ tile, $8*8 \rightarrow 16$ bits \\ Single x-pointer is used to access samples from column, separated by N bytes \\ 10 vector registers used}\label{GEMM_2xVLEN_b_m2}
        }
    \end{center}

\begin{center}
\small
\texttt{
        \begin{tabular}{ |l | l | c | c | c  |}
        \hline
Cycle & Instruction issued  & Unit  &       VMAC:M      &  VLOAD:L \\
        \hline
0 & vle8.v         v8,(x28)               &     VLOAD & M:v1:0,v1:1  &  L:v8:0 \\
1 & c.addi         x28,0x10               &    Scalar & M:v2:0,v2:1  &  L:v8:1 \\
2 & lb             x7,(x13)               &    Scalar & M:v3:0,v3:1  &   .\\
3 & vwmacc.vx      v0,x7,v8              &      VMAC & .             &   .  \\
4 & c.addi         x13,0x1                &    Scalar & .            &       .  \\
5 & lb             x8,(x5)                &    Scalar & .            &       .  \\
6 & c.addi         x5,0x1                 &    Scalar & .            &       .  \\
7 & vwmacc.vx      v4,x8,v8              &      VMAC &      .       &       .  \\
8 & bne            x13,x15,0xffffffe6     &    Scalar & M:v0:0,v0:1  &       .  \\
       \hline
        \end{tabular}
\captionof{table}{Matrix multiplication kernel, $2\times VLEN$ tile, $8*8 \rightarrow 16$ bits \\ 4 vector registers used}\label{GEMM_2xVLEN_m1}
        }
    \end{center}

\begin{center}
\small
\texttt{
        \begin{tabular}{ |l | l | c | c | c  |}
        \hline
Cycle & Instruction issued  & Unit  &       VMAC:M      &  VLOAD:L \\
        \hline
0  & vle8.v         v16,(x28)               &    VLOAD & M:v5:0,v5:1     &  L:v16:0      \\
1  & addi           x28,x28,0x20            &   Scalar & M:v6:0,v6:1     &  L:v16:1     \\
2  & lh             x7,(x13)                &   Scalar & M:v7:0,v7:1     &  L:v17:0     \\
3  & -              -                       &        - & M:v8:0,v8:1     &  L:v17:1     \\
4  & -              -                       &        - & M:v9:0,v9:1     &  L:v18:0     \\
5  & -              -                       &        - & M:v10:0,v10:1   &    L:v18:1   \\
6  & vwmacc.vx      v0,x7,v16               &     VMAC & M:v11:0,v11:1   &    L:v19:0   \\
7  & c.addi         x13,0x2                 &   Scalar & M:v12:0,v12:1   &    L:v19:1   \\
8  & lh             x8,(x5)                 &   Scalar & M:v13:0,v13:1   &         .    \\
9  & c.addi         x5,0x2                  &   Scalar & M:v14:0,v14:1   &         .    \\
10  & -              -                       &        - & M:v15:0,v15:1   &         .    \\
11  & -              -                       &        - & M:v0:0,v0:1     &       .      \\
12  & -              -                       &        - & M:v1:0,v1:1     &       .      \\
13  & -              -                       &        - & M:v2:0,v2:1     &       .      \\
14  & vwmacc.vx      v8,x8,v16               &     VMAC & M:v3:0,v3:1     &       .      \\
15  & bne            x13,x15,0xffffffe4      &   Scalar & M:v4:0,v4:1     &       .      \\
       \hline
        \end{tabular}
\captionof{table}{Matrix multiplication kernel, $2 \times VLEN\cdot 4 $ tile, $8*8 \rightarrow 16$ bits \\ 20 vector registers used}\label{GEMM_2xVLEN_m4}
}
    \end{center}

Summarizing, while RVV 1.0 32 vector registers provide ample margin, 10 vector registers is almost as good for this kernel. Also note that $2\times VLEN \cdot 4$ tile size might be impractical for some applications: in small embedded applications one is often dealing with smaller matrices. And for e.g. 8-bit data $2\times VLEN \cdot 4$ (2x32 elements) might be too wide a tile.

\subsection { Vector Accumulation } 

Accumulation appears in many applications, for instance if one needs to compute the sum or average of an array of data samples. The standard approach is to accumulate data in a vector and then do a reduction sum of the vector. For sizeable vectors most cycles are spent in the accumulation loop, so we focus on it.

Assuming the processor is capable of writing double-wide results to a vector register each cycle, we have the kernel shown in Table \ref{accum_m2}, with 100\% ALU and load bandwidth utilization.

\begin{center}
\small
\texttt{
        \begin{tabular}{ |l | l | c | c | c  |}
        \hline
Cycle & Instruction issued                & Unit      &  ALU:A                & VLOAD:L \\
        \hline
0 & vl4re16.v      v0,(a0)                &     VLOAD &        A:v2:0,v2:1    &   L:v0:0 \\
1 & vwadd.wv       v2,v2,v0               &       ALU &        A:v3:0,v3:1    &   L:v0:1 \\
2 & addi           a0,a0,16               &    Scalar &        A:v4:0,v4:1  &   L:v1:0 \\
3 & bne            a5,a0,pc - 16          &    Scalar &        A:v5:0,v5:1  &   L:v1:1 \\
       \hline
        \end{tabular}
\captionof{table}{Vector accumulation, $VLEN\cdot 2$ tile (LMUL=2), $16*16 \rightarrow 32$ bits \\ 6 vector registers used}\label{accum_m2}
        }
    \end{center}

\subsection { Dot Product }
Dot product is a building block of many linear algebra and DSP kernels. In the vector*vector form it is load-bound, requiring two vector loads per one vector MAC, so if vector loads are only DLEN-wide we can at best have 50\% MAC utilization. LMUL=2 is enough to get to full utilization of the load unit, as shown in Table \ref{dot_prod_m2}

\begin{center}
\small
\texttt{
        \begin{tabular}{ |l | l | c | c | c  |}
        \hline
Cycle & Instruction issued                & Unit      &  MAC:M        & VLOAD:L \\
        \hline
0 & vle16.v        v0,(x10)               &     VLOAD &      .        &   L:v0:0   \\
1 & c.addi         x10,0x10               &    Scalar &      .        &   L:v0:1   \\
2 & -              -                      &         - &      .        &   L:v1:0   \\
3 & -              -                      &         - & M:v4:0,v4:1   &   L:v1:1   \\
4 & vle16.v        v2,(x11)               &     VLOAD & M:v5:0,v5:1   &   L:v2:0   \\
5 & c.addi         x11,0x10               &    Scalar & M:v6:0,v6:1   &   L:v2:1   \\
6 & vwmacc.vv      v4,v0,v2               &      VMAC & M:v7:0,v7:1   &   L:v3:0   \\
7 & bne            x15,x10,0xfffffff0     &    Scalar &      .        &   L:v3:1   \\
       \hline
        \end{tabular}
\captionof{table}{Dot product, $VLEN\cdot 2$ tile (LMUL=2), $16*16 \rightarrow 32$ bits \\ 8 vector registers used}\label{dot_prod_m2}
}
    \end{center}

\subsection { Matrix by Vector Multiplication }

Matrix by vector multiplication is a common kernel, used for AI/inference (e.g., dense layers, perceptron-like) and DSP (e.g., Finite Impulse Response filter). For the processor considered this is both a load- and MAC-bound kernel, and best utilization is with LMUL=4, with 4 vector registers used for input and 8 vector registers used for output, as shown in Table \ref{gemv_m4}: 9 cycles, of which MAC is active in 8 cycles, so 8/9 = 89\%

\begin{center}
\small
\texttt{
        \begin{tabular}{ |l | l | c | c | c  |}
        \hline
Cycle &  Instruction issued & Unit & MAC:M & LOAD:L \\
        \hline
0 & lbu            a2,0(a5)              &     Scalar & M:v10:0,v10:1 &     a2   \\
1 & vle8.v         v0,(a4)               &      VLOAD & M:v9:0,v9:1   &   L:v0:0  \\
2 & c.addi         a5,1                  &     Scalar & M:v11:0,v11:1 &     L:v0:1  \\
3 & addi           a4,a4,128             &     Scalar & M:v4:0,v4:1   &   L:v1:0  \\
4 &  -              -                     &          - & M:v5:0,v5:1   &   L:v1:1  \\
5 & -              -                     &          - & M:v6:0,v6:1   &   L:v2:0  \\
6 & vwmacc.vx      v4,v0,a2              &       VMAC & M:v7:0,v7:1   &   L:v2:1  \\
7 & bne            s0,a5,pc - 18         &     Scalar & M:v8:0,v8:1   &   L:v3:0  \\
8 &  -              -                    &            &               &   L:v3:1  \\
       \hline
        \end{tabular}
\captionof{table}{Matrix by vector multiplication, $ VLEN \cdot 4 $ tile (LMUL=4), $8*8 \rightarrow 16 bits $ \\ 12 vector registers used}\label{gemv_m4}
        }
    \end{center}

\subsection{Results Summary}
Table \ref{table_summary} summarizes kernels for a few tiling sizes, LMUL factor used, required vector register number and resulting MAC utilization.

An increase of the number of vector registers beyond the indicated numbers does not improve MAC utilization further. All considered microkernels are either MAC-bound or load/store bandwidth bound. 

\begin{center}
\small
        \begin{tabular}{ | l | c | c | p{2cm} | p{2cm} | l |}
        \hline
 Kernel & Data Size & LMUL  & Vector \newline registers used & MAC \newline Utilization, \%  & Comment  \\
         \hline
 Matrix*Matrix &   $2\times (VLEN\cdot 2)$ & 2  & 10 & 89 to 100  & Depends on size (+addi) \\
 Matrix*Matrix &   $2\times VLEN$ & 1  & 4 & 44 to 50  &  tile too small for efficiency\\
 Accumulation & $VLEN\cdot 2$ & 2 & 6 & 100 & \\
 Dot product & $VLEN\cdot 2$ & 2 & 8 & 50 & Load-limited \\
 Matrix*Vector & $VLEN\cdot 4$ & 4 & 12 & 89 & \\
          \hline
        \end{tabular}      
\captionof{table}{Results summary for kernels, number of vector registers used and MAC utilization}\label{table_summary}        
\end{center}

\section{ Conclusions }
Our analysis shows that important kernels can be efficiently implemented with fewer than 32 vector registers, the number required by the RVV 1.0 standard. This applies to vector lengths as small as 64 bits. All kernels fit into 16 vector registers. Some fit into 8 vector registers and all fit into 8 registers with lower utilization.  

Configurations with a reduced number of vector registers enable smaller processor designs. This approach does not require a completely new programming model. Although the resulting processor does not have binary code compatibility with standard RVV, compiling for it only requires parameterization of the vector register file size in the compiler. Overall optimization approaches are still the same, allowing to have a unified source code base. This appears to be a good choice for deeply embedded designs, for which unconstrained binary code compatibility is not a hard requirement.

Our result suggests an ISA extension to reduce the number of vector registers to 16 or 8 as part of the RVV standard.

$\,$

$\,$

\bibliographystyle{plain}
\bibliography{RVV_Reduced_Vregs}



\end{document}